\documentclass[twocolumn,showpacs,amsmath,amssymb,prl]{revtex4}

\usepackage{graphicx}
\usepackage{dcolumn}
\usepackage{bm}

\begin{document}


\title{Improved Test of Lorentz Invariance in Electrodynamics}

\author{Peter Wolf}
 \altaffiliation[On leave from ]{Bureau International des Poids et Mesures, Pavillon de Breteuil, 92312 S\`evres Cedex, France.}
\author{S\'ebastien Bize}%
\author{Andr\'e Clairon}%
\author{Giorgio Santarelli}%
\affiliation{BNM-SYRTE, Observatoire de Paris, 61 Av. de l'Observatoire, 75014 Paris, France}%

\author{Michael E. Tobar}
\author{Andr\'e N. Luiten}
\affiliation{University of Western Australia, School of Physics, Nedlands 6907 WA, Australia}%

\date{\today}

\begin{abstract}
We report new results of a test of Lorentz invariance based on the comparison of a cryogenic sapphire microwave resonator and a hydrogen maser. The experimental results are shown together with an extensive analysis of systematic effects. Previously, this experiment has set the most stringent constraint on Kennedy-Thorndike type violations of Lorentz invariance. In this work we present new data and interpret our results in the general Lorentz violating extension of the standard model of particle physics (SME). Within the photon sector of the SME, our experiment is sensitive to seven SME parameters. We marginally improve present limits on four of these, and by a factor 7 to 10 on the other three.
\end{abstract}

\pacs{03.30.+p, 06.30.Ft, 12.60.-i, 11.30.Cp, 84.40.-x}
\maketitle

Lorentz Invariance (LI) is the fundamental postulate of Special Relativity and therefore at the heart of all accepted theories of physics. The central importance of this postulate in modern physics has motivated tremendous work to experimentally test LI with ever increasing precision. Additionally, many unification theories (e.g. string theory or loop gravity) are expected to violate LI at some level \cite{KostoSam,Damour1,Gambini}, which further motivates experimental searches for such violations.

Numerous test theories that allow the modeling and interpretation of experiments that test LI have been developed \cite{Robertson,MaS,LightLee,Ni,Kosto1,KM}. The kinematical frameworks (RMS) of \cite{Robertson, MaS} postulate a simple parametrisation of the Lorentz transformations with experiments setting limits on the deviation of those parameters from their special relativistic values. Owing to their simplicity they have been widely used to model and interpret many experiments that test LI, including our previous publications \cite{Wolf,WolfGRG}. More recently, a general Lorentz violating extension of the standard model of particle physics (SME) has been developed \cite{Kosto1} whose Lagrangian includes all parametrised Lorentz violating terms that can be formed from known fields. Over the last years considerable experimental effort has gone into SME tests \cite{KostoB,Lipa,Muller}, and the present paper reports our first results in the SME.

In the photon sector of the SME \cite{KM}, Lorentz violating terms are parametrised by 19 independent components of a tensor $(k_F)_{\kappa\lambda\mu\nu}$ (greek indices run from 0 to 3), which are in general grouped into three traceless and symmetric $3\times 3$ matrices ($\tilde{\kappa}_{e+}$, $\tilde{\kappa}_{o-}$, and $\tilde{\kappa}_{e-}$), one antisymmetric one ($\tilde{\kappa}_{o+}$) and one additional scalar, which all vanish when LI is satisfied. The 10 independent components of $\tilde{\kappa}_{e+}$ and $\tilde{\kappa}_{o-}$ are constrained by astronomical measurements to $< 2\times 10^{-32}$. Seven components of $\tilde{\kappa}_{e-}$ and $\tilde{\kappa}_{o+}$ have been constrained in an optical cavity experiment \cite{Muller} at the $10^{-15}$ and $10^{-11}$ level respectively, while the remaining two prameters are unconstrained for the time being. Our experiment is sensitive to the same seven components of $\tilde{\kappa}_{e-}$ and $\tilde{\kappa}_{o+}$ as \cite{Muller}. It marginally improves present limits on four of these, with significant (factor 7 to 10) improvement on the other three.

As already described in \cite{Wolf,WolfGRG}, our experiment consists of a cryogenic sapphire oscillator (CSO) operating in a whispering gallery mode with a resonance frequency of 11.932 GHz which is compared to a commercial (Datum Inc.) active hydrogen maser. A detailed description can be found in \cite{Chang,Mann,ChangTh} with specific details in the context of LI tests in \cite{WolfGRG}.

A standing wave is set up around the circumference of the cylindrical sapphire resonator with the dominant electric and magnetic fields in the radial and vertical directions, corresponding to a propagation (Poynting) vector around the circumference. In the photon sector of the SME the resonant frequency of an electro-magnetic cavity is subject to a Lorentz violating perturbation which can be expressed, to first order, as an integral over the non-perturbed e-m fields (equ. (34) of \cite{KM}). The change of orientation of the fields due to the rotation and orbital motion of the Earth then provides a time varying modulation of that perturbed frequency, which is searched for in the experiment. As shown in \cite{KL} the frequency of the H-maser is not affected to first order (because it operates on $m_F=0$ states) and \cite{Muller2} shows that the perturbation of the frequency due to the modification of the sapphire crystal structure (and hence the cavity size) is negligible with respect to the direct perturbation of the e-m fields.

With the above assumptions the perturbed frequency difference $\nu$ between the CSO and the H-maser can be expressed in the form (see \cite{WolfGRG} for a detailed derivation)
\begin{equation}
\frac{\nu-\nu_0}{\nu_0} = \sum_i C_i {\rm cos}(\omega_i T_\oplus + \varphi_i) + S_i {\rm sin}(\omega_i T_\oplus + \varphi_i)\label{nuSME}
\end{equation}
where $\nu_0$ is the unperturbed frequency difference, the sum is over the six frequencies $\omega_i$ of Tab.\ref{Tab3}, the coefficients $C_i$ and $S_i$ are functions of the Lorentz violating tensors $\tilde{\kappa}_{e-}$ and $\tilde{\kappa}_{o+}$ (see Tab.\ref{Tab3}), $T_\oplus = 0$ on December 17, 2001, 18:05:16 UTC, $\varphi_{\omega_\oplus}=\varphi_{2\omega_\oplus}=0$ and $\varphi_{(\omega_\oplus\pm\Omega_\oplus)}=\varphi_{(2\omega_\oplus\pm\Omega_\oplus)}=\pm 4.682$ rad.

\begin{table*}
\caption{\label{Tab3} Coefficients $C_i$ and $S_i$ in (1) for the six frequencies $\omega_i$ of interest and their relation to the components of the SME parameters $\tilde{\kappa}_{e-}$ and $\tilde{\kappa}_{o+}$, with $\omega_\oplus$ and $\Omega_\oplus$ the angular frequencies of the Earth's sidereal rotation and orbital motion. The measured values (in $10^{-16}$) are shown together with the statistical (first bracket) and systematic (second bracket) uncertainties.}
\begin{ruledtabular}
\begin{tabular}{ccccc}
$\omega_i$ &\multicolumn{2}{c}{$C_i$}&\multicolumn{2}{c}{$S_i$}\\
\hline
$\omega_\oplus - \Omega_\oplus$ & $(-8.6\times 10^{-6})\tilde{\kappa}_{o+}^{YZ}$ & $-6.9(4.2)(4.5)$ & $(8.6\times 10^{-6})\tilde{\kappa}_{o+}^{XZ}-(4.2\times 10^{-5})\tilde{\kappa}_{o+}^{XY}$ & $6.7(4.2)(4.5)$\\
$\omega_\oplus$ & $-0.44\tilde{\kappa}_{e-}^{XZ}+(1.1\times 10^{-6})\tilde{\kappa}_{o+}^{XZ}$ & $14(4.2)(4.2)$ & $-0.44\tilde{\kappa}_{e-}^{YZ}+(1.1\times 10^{-6})\tilde{\kappa}_{o+}^{YZ}$ & $2.4(4.2)(4.2)$\\
$\omega_\oplus + \Omega_\oplus$ & $(-8.6\times 10^{-6})\tilde{\kappa}_{o+}^{YZ}$ & $-6.0(4.2)(4.2)$ & $(8.6\times 10^{-6})\tilde{\kappa}_{o+}^{XZ}+(1.8\times 10^{-6})\tilde{\kappa}_{o+}^{XY}$ & $2.7(4.2)(4.2)$\\
$2\omega_\oplus - \Omega_\oplus$ & $(-1.8\times 10^{-5})\tilde{\kappa}_{o+}^{XZ}$ & $3.7(2.4)(3.7)$ & $(-1.8\times 10^{-5})\tilde{\kappa}_{o+}^{YZ}$ & $-2.9(2.4)(3.7)$\\
$2\omega_\oplus$ & $-0.10(\tilde{\kappa}_{e-}^{XX}-\tilde{\kappa}_{e-}^{YY})$ & $3.1(2.4)(3.7)$ & $-0.19\tilde{\kappa}_{e-}^{XY}$ & $11(2.4)(3.7)$\\
$2\omega_\oplus + \Omega_\oplus$ & $(7.8\times 10^{-7})\tilde{\kappa}_{o+}^{XZ}$ & $0.0(2.4)(3.7)$ & $(7.8\times 10^{-7})\tilde{\kappa}_{o+}^{YZ}$ & $-1.2(2.4)(3.7)$\\
\end{tabular}
\end{ruledtabular}
\end{table*}

Our previously published results \cite{Wolf,WolfGRG} are based on data sets taken between Nov. 2001 and Sep. 2003. All of the data in \cite{Wolf} (except the last data set) were taken before implementation of permanent temperature control of the local environment. As a result the uncertainties in \cite{Wolf} were dominated by the systematic effects from temperature variations. In \cite{WolfGRG} we have used only data that were permanently temperature controlled yielding an improvement on \cite{Wolf} by about a factor 2 for the RMS parameters of \cite{Robertson,MaS}. However, those data were insufficient to decorrelate all 7 SME parameters.

In order to do so we have extended the data to 20 data sets in total, spanning Sept. 2002 to Jan. 2004, of differing lengths (5 to 20 days, 222 days in total). The sampling time for all data sets was $100$ s. Fig. 1. shows the complete data and the power spectral density (PSD) of the longest data set after removal of an offset and a rate (natural frequency drift, typically $\approx 1.6 \times 10^{-18}$ s$^{-1}$) for each data set.

For the statistical analysis we first average the data to 2500 s sampling time and then simultaneously fit the 20 rates and offsets and the 12 parameters $C_i$ and $S_i$ of (\ref{nuSME}) to the complete data using two statistical methods, weighted least squares (WLS), which allows one to account for non-white noise processes (cf. \cite{Wolf}), and individual periods (IP) as used in \cite{Muller}. The two methods give similar results for the parameters (within the uncertainties) but differ in the estimated uncertainties (the IP uncertainties are a factor $\approx 1.2$ larger). Because IP discards a significant amount of data (about 10\% in our case) we consider WLS the more realistic method and retain those results as the statistical uncertainties shown in Tab. \ref{Tab3}. We note that we now have sufficient data to decorrelate all 12 parameters ($C_i$, $S_i$) i.e. the WLS correlation coefficients between any two parameters or between any parameter and the fitted offsets and rates are all less than 0.20.

To investigate the distributions of our results we fit the coefficients $C_i$ and $S_i$ to each one of the 20 data sets individually with the results at the sidereal and semi-sidereal frequencies $\omega_\oplus$ and $2\omega_\oplus$ shown in Fig. 2. If a genuine effect at those frequencies was present we would expect correlated phases of the individual points in Fig. 2., but this does not seem to be supported by the data. A distribution of the phases may result from an effect at a neighbouring frequency, in particular the diurnal and semi-diurnal frequencies $\omega_\oplus - \Omega_\oplus$ and $2(\omega_\oplus - \Omega_\oplus)$ at which we would expect systematic effects to play an important role. 
Fig. 3. shows the amplitudes $A_\omega = \sqrt{C_\omega^2+S_\omega^2}$ resulting from least squares fits for a range of frequencies, $\omega$, around the frequencies of interest. We note that the fitted amplitudes at $\omega_\oplus - \Omega_\oplus$ and $2(\omega_\oplus - \Omega_\oplus)$ are substantially smaller than those at $\omega_\oplus$ and $2\omega_\oplus$ and therefore unlikely to contribute to the distribution of the points in Fig. 2.

Systematic effects at the frequencies $\omega_i$ could mask a putative Lorentz violating signal in our experiment and need to be investigated in order to be able to confirm such a signal or to exclude it within realistic limits. We have extensively studied all systematic effects arising from environmental factors that might affect our experiment. The resulting estimated contributions at the two central frequencies $\omega_\oplus$, $2\omega_\oplus$ and at the diurnal frequency $\omega_\oplus - \Omega_\oplus$ are summarised in Tab.\ref{Tab2}. The contributions at $\omega_\oplus + \Omega_\oplus$ and $2\omega_\oplus \pm \Omega_\oplus$ are not shown as they are identical to those at $\omega_\oplus$ and $2\omega_\oplus$ respectively.

\begin{table}
\caption{\label{Tab2}  Contributions from systematic effects to the amplitudes $A_i$ (parts in $10^{16}$) at three frequencies $\omega_i$.}
\begin{ruledtabular}
\begin{tabular}{lccc}
Effect & $\omega_\oplus - \Omega_\oplus$ & $\omega_\oplus$ & $2\omega_\oplus$ \\
\hline
H-maser & $< 5$ & $< 5$ & $< 5$\\
Tilt & 3 & 3 & 1\\
Gravity & 0.3 & 0.3 & 0.3\\
B-field & $< 0.1$ & $< 0.1$ & $< 0.1$\\
Temperature & $< 1$ & $< 1$ & $< 1$\\
Atm. Pressure & 2.3 & 0.3 & 0.4\\
\hline
{\bf Total} & {\bf 6.4} & {\bf 5.9} & {\bf 5.2}\\
\end{tabular}
\end{ruledtabular}
\end{table}

We have compared the Hydrogen-maser (HM) used as our frequency reference to our highly stable and accurate Cs fountain clocks (FO2 and FOM). For example, the amplitudes at $\omega_\oplus$ and $2\omega_\oplus$ of the HM-FOM relative frequency difference over June-July 2003 were $A_{\omega_\oplus}=(4.8\pm 4.7)\times 10^{-16}$ and $A_{2\omega_\oplus}=(4.3\pm 4.7)\times 10^{-16}$. This indicates that any environmental effects on the HM at those frequencies should be below 5 parts in $10^{16}$ in amplitude. This is in good agreement with studies on similar HMs carried out in \cite{Parker} that limited environmental effects to $<$ 3 to 4 parts in $10^{16}$.

To estimate the tilt sensitivity we have intentionally tilted the oscillator by $\approx$ 5 mrad off its average position which led to relative frequency variations of $\approx 3 \times 10^{-13}$ from which we deduce a tilt sensitivity of $\approx 6 \times 10^{-17} \mu$rad$^{-1}$. This is in good agreement with similar measurements in \cite{ChangTh} that obtained sensitivities of $\approx 4 \times 10^{-17} \mu$rad$^{-1}$. Measured tilt variations in the lab at diurnal and semi-diurnal periods show amplitudes of 4.6 $\mu$rad and 1.6 $\mu$rad respectively which leads to frequency variations that do not exceed $3 \times 10^{-16}$ and $1 \times 10^{-16}$ respectively.

From the measurements of tilt sensitivity one can deduce the sensitivity to gravity variations (cf. \cite{ChangTh}), which in our case lead to a sensitivity of $\approx 3 \times 10^{-10} g^{-1}$. Tidal gravity variations can reach $\approx 10^{-7} g$ from which we obtain a maximum effect of $3 \times 10^{-17}$, one order of magnitude below the effect from tilt variations.

Variations of the ambient magnetic field in our lab. are dominated by the passage of the Paris Metro, showing a strong periodicity ("quiet" periods from 1 am to 5 am). The corresponding diurnal and semi-diurnal amplitudes are $1.7 \times 10^{-4}$ G and $3.4 \times 10^{-4}$ G respectively for the vertical field component and about 10 times less for the horizontal one. To determine the magnetic sensitivity of the CSO we have applied a sinusoidal vertical field of 0.1 G amplitude with a 200 s period. Comparing the CSO frequency to the FO2 Cs-fountain we see a clear sinusoidal signal (S/N $> 2$) at the same period with an amplitude of $7.2 \times 10^{-16}$, which leads to a sensitivity of $\approx 7 \times 10^{-15}$ G$^{-1}$. Assuming a linear dependence (there is no magnetic shielding that could lead to non-linear effects) we obtain effects of only a few parts in $10^{-18}$.

Late 2002 we implemented an active temperature stabilization inside an isolated volume ($\approx 15 {\rm m}^3$) that includes the CSO and all the associated electronics. The temperature is measured continously in two fixed locations (behind the electronics rack and on top of the dewar). For the best data sets the measured temperature variations do not exceed 0.02/0.01 K in amplitude for the diurnal and semi-diurnal components. A least squares fit to all our temperature data (taken simultaneously with our frequency measurements) yields amplitudes of $A_{\omega_\oplus}=0.020$ K and $A_{2\omega_\oplus}=0.018$ K with similar values at the other frequencies $\omega_i$ of interest, including the diurnal one ($A_{\omega_\oplus - \Omega_\oplus}=0.022$ K). Inducing a strong sinusoidal temperature variation ($\approx 0.5$ K amplitude at 12 h period) leads to no clearly visible effect on the CSO frequency. Taking the noise level around the 12 h period as the maximum effect we obtain a sensitivity of $< 4 \times 10^{-15}$ per K. Using this estimate we obtain effects of $< 1 \times 10^{-16}$ at all frequencies $\omega_i$.

Finally we have investigated the sensitivity of the CSO to atmospheric pressure variations. To do so we control the pressure inside the dewar using a variable valve mounted on the He-gas exhaust. During normal operation the valve is open and the CSO operates at ambient atmospheric pressure. For the sensitivity determination we have induced a sinusoidal pressure variation ($\approx 14$ mbar amplitude at 12 h period), which resulted in a clearly visible effect on the CSO frequency corresponding to a sensitivity of $\approx 6.5 \times 10^{-16}$ mbar$^{-1}$. We have checked that the sensitivity is not significantly affected when changing the amplitude of the induced pressure variation by a factor 3. A least squares fit to atmospheric pressure data (taken simultaneously with our frequency measurements) yields amplitudes of $A_{\omega_\oplus}=0.045$ mbar and $A_{2\omega_\oplus}=0.054$ mbar with similar values at the other frequencies $\omega_i$ of interest, except the diurnal one for which $A_{\omega_\oplus - \Omega_\oplus}=0.36$ mbar. The resulting effects on the CSO frequency are given in Tab.\ref{Tab2}.

Our final results for the 7 components of $\tilde{\kappa}_{e-}$ and $\tilde{\kappa}_{o+}$ are obtained from a least squares fit to the 12 measured coefficients of Tab.\ref{Tab3}. They are summarised and compared to the results of \cite{Muller} in Tab.\ref{Tab4}.

\begin{table}
\caption{\label{Tab4}Results for the components of the SME Lorentz violation parameters $\tilde{\kappa}_{e-}$ (in $10^{-15}$) and $\tilde{\kappa}_{o+}$ (in $10^{-11}$).}
\begin{ruledtabular}
\begin{tabular}{ccccc}
& $\tilde{\kappa}_{e-}^{XY}$ & $\tilde{\kappa}_{e-}^{XZ}$ & $\tilde{\kappa}_{e-}^{YZ}$ & $(\tilde{\kappa}_{e-}^{XX}-\tilde{\kappa}_{e-}^{YY})$ \\
\hline
from \cite{Muller} & 1.7(2.6) & -6.3(12.4) & 3.6(9.0) & 8.9(4.9)\\
this work & -5.7(2.3) & -3.2(1.3) & -0.5(1.3) & -3.2(4.6)\\
\hline \hline
& $\tilde{\kappa}_{o+}^{XY}$ & $\tilde{\kappa}_{o+}^{XZ}$ & $\tilde{\kappa}_{o+}^{YZ}$ &\\
\hline
from \cite{Muller} & 14(14) & -1.2(2.6) & 0.1(2.7) &\\
this work & -1.8(1.5) & -1.4(2.3) & 2.7(2.2) &
\end{tabular}
\end{ruledtabular}
\end{table}

We note that our results for $\tilde{\kappa}_{e-}^{XY}$ and $\tilde{\kappa}_{e-}^{XZ}$ are significant at about $2\sigma$, while those of \cite{Muller} are significant at about the same level for $(\tilde{\kappa}_{e-}^{XX}-\tilde{\kappa}_{e-}^{YY})$. The two experiments give compatible results for $\tilde{\kappa}_{e-}^{XZ}$ (within the $1\sigma$ uncertainties) but not for the other two parameters, so the measured values of those are unlikely to come from a common source. Another indication for a non-genuine effect comes from Figs. 2. and 3., as we would expect any genuine effect to show an approximately coherent phase for the individual data sets in Fig. 2. and to display more prominent peaks in Fig. 3.

In conclusion, we have not seen any Lorentz violating effects in the general framework of the SME, and set limits on 7 parameters of the SME photon sector (cf. Tab. \ref{Tab4}) which are up to an order of magnitude more stringent than those obtained from previous experiments \cite{Muller}. Two of the parameters are significant (at $\approx 2\sigma$). We believe that this is most likely a statistical coincidence or a neglected systematic effect. To verify this, our experiment is continuing and new, more precise experiments are under way \cite{Mike}.

\begin{acknowledgments}
Helpful discussions with Alan V. Kosteleck\'y, and partial funding by the Australian Research Council are greatfully acknowledged.
\end{acknowledgments}

\vspace{1cm}
\noindent{\bf Figure captions}

\noindent Fig. 1: Relative frequency difference between the CSO and the H-maser after removal of a linear fit per data set. Complete data (inset) and PSD of the longest data set.

\vspace{3mm}

\noindent Fig. 2: Fitted sine and cosine amplitudes at $\omega_\oplus$ and $2\omega_\oplus$ for each data set (blue squares) and the complete data (red diamonds, with statistical errors). For clarity the error bars of the individual data sets have been omitted.

\vspace{3mm}

\noindent Fig. 3: Fitted Amplitudes $A_\omega$ for a range of frequencies around the six frequencies $\omega_i$ of interest (indicated by arrows).

\begin{figure*}
\includegraphics{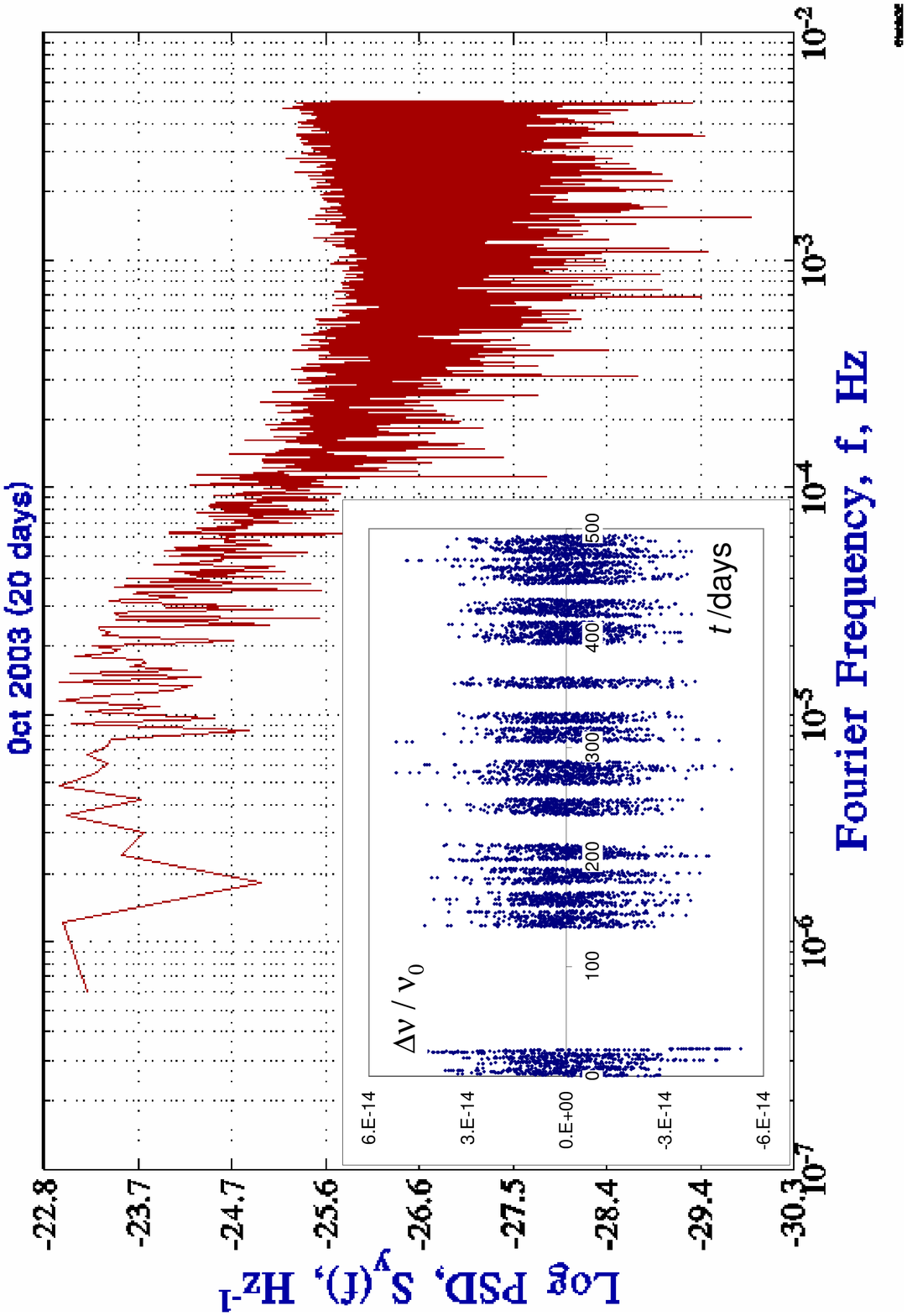}
\end{figure*}

\begin{figure*}
\includegraphics{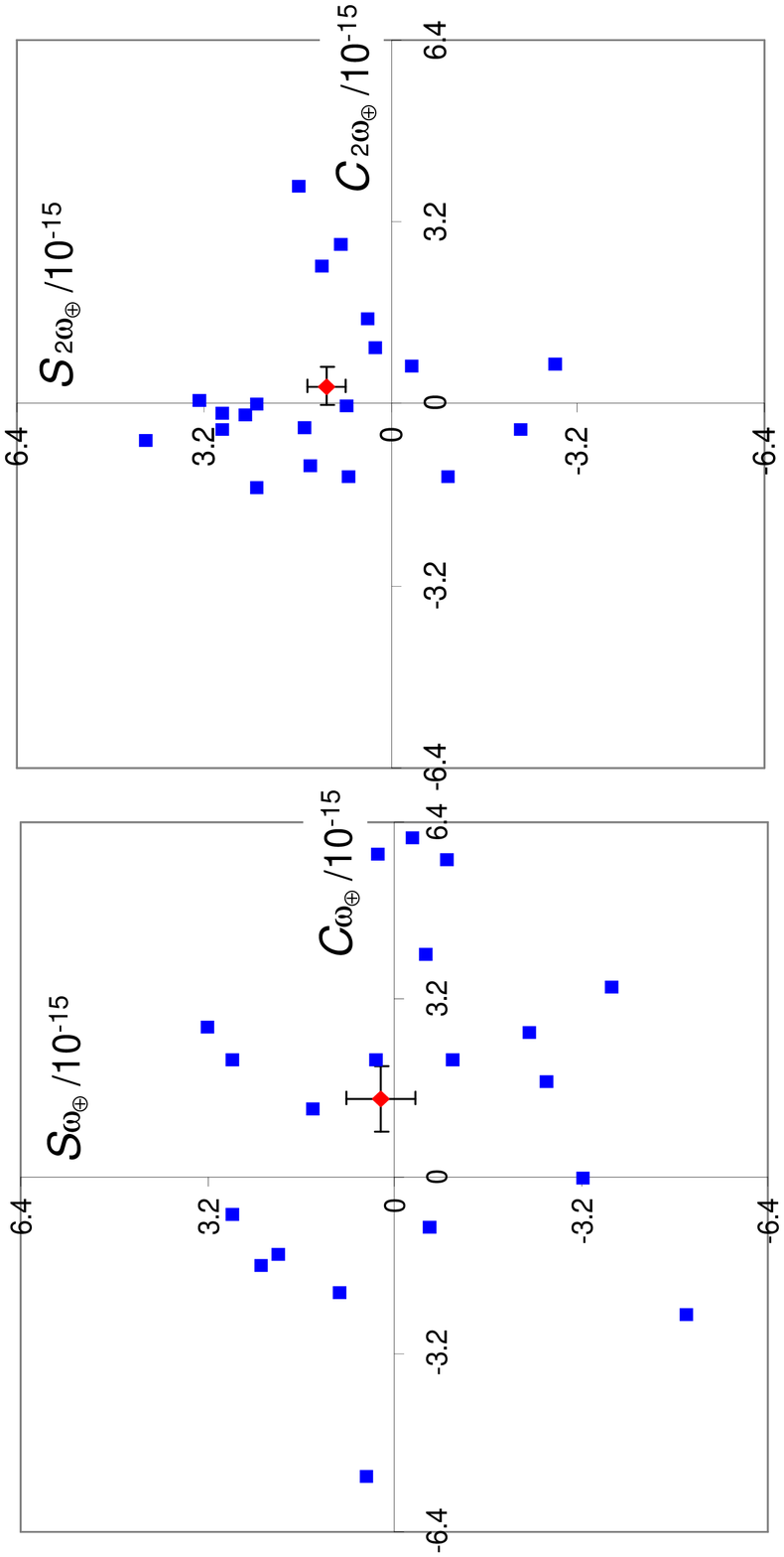}
\end{figure*}

\begin{figure*}
\includegraphics{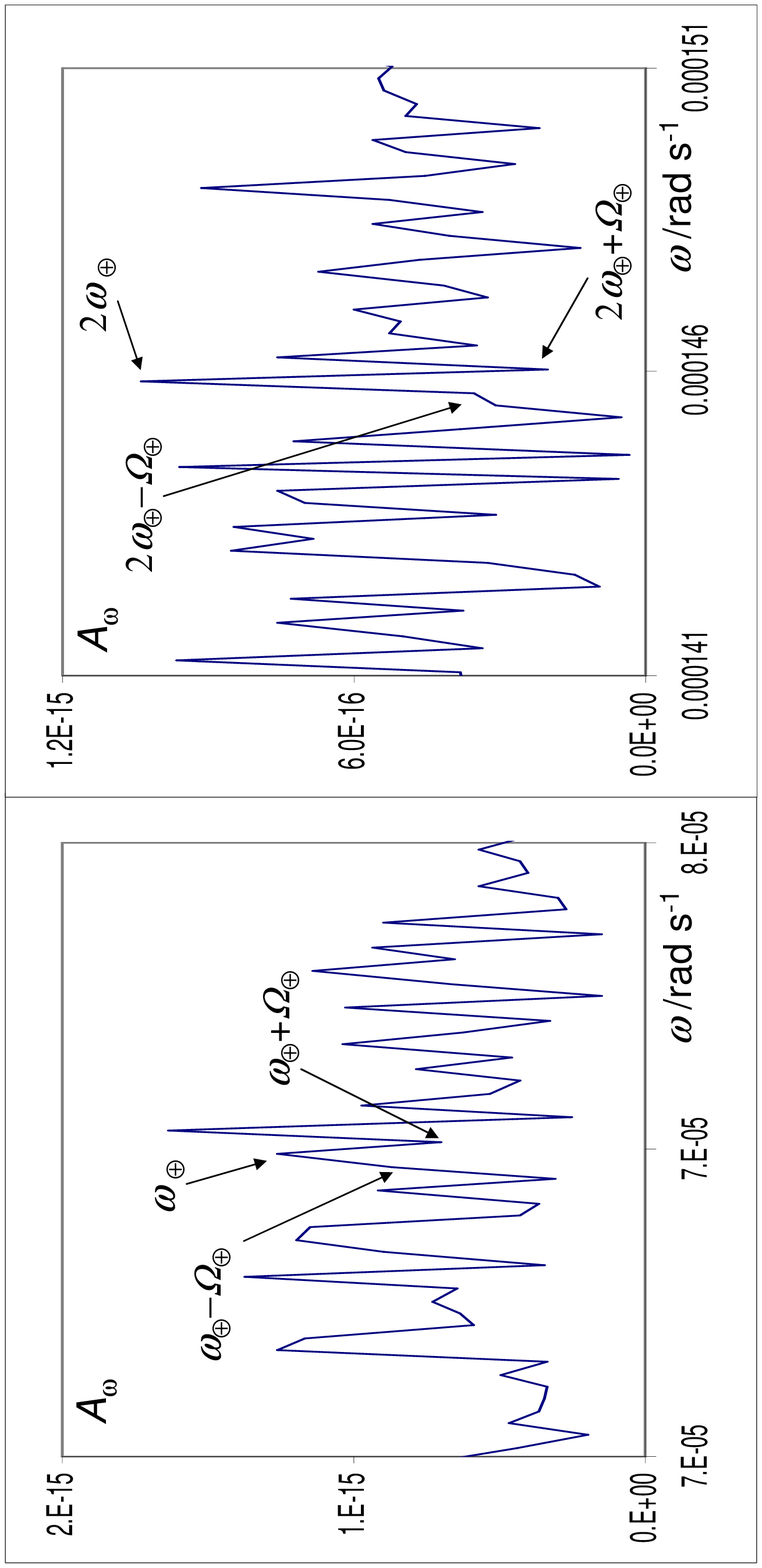}
\end{figure*}

\end{document}